\documentclass{aa}
\usepackage[varg]{txfonts}
\usepackage{natbib}
\usepackage{url}
\usepackage{hyperref}
\usepackage{xcolor}
\usepackage{ulem}
\usepackage{orcidlink}
\usepackage{xspace}

\usepackage{upgreek}
\usepackage{graphicx}   
\usepackage{multirow}
\usepackage{subcaption} 

\newcommand{\src}{{1A~1118$-$61}\xspace}
\newcommand{\nustar}{\textit{NuSTAR}\xspace}

\newcommand{\hxmt}{\textit{Insight-HXMT}\xspace}

\newcommand{\srg}{\textit{SRG}\xspace}

\def\lum{erg\,s$^{-1}$\xspace}

\begin{document}

\title{The rise and fall of \src: First evidence for frequency hysteresis of quasi-periodic oscillations in an X-ray pulsar}

\titlerunning{Hysteresis of quasi-periodic oscillations in \src}

\authorrunning{Salganik, A., et al.}
\author{Alexander~Salganik\inst{\ref{in:UTU}}\orcidlink{0000-0003-2609-8838} \corrauth{alsalganik@gmail.com} , 
Sergey~S.~Tsygankov\inst{\ref{in:UTU},\ref{in:IHEP}}\orcidlink{0000-0002-9679-0793} 
\and    Sergey~V.~Molkov\inst{\ref{in:IKI}}\orcidlink{0000-0002-5983-5788}
\and Hua Xiao\inst{\ref{in:UTU}} \orcidlink{0009-0004-1288-4912}
\and  Lingda~Kong\inst{\ref{in:NU},\ref{in:IAAT}} \orcidlink{0000-0003-3188-9079}
\and
QingChang~Zhao\inst{\ref{in:IHEP}, \ref{in:UCAS}}\orcidlink{0000-0001-9893-8248}
\and 
Long Ji\inst{\ref{in:SYS}}\orcidlink{0000-0001-9599-7285}
\and  Alexander A. Mushtukov\inst{\ref{in:UCL},\ref{in:OXF}}\orcidlink{0000-0003-2306-419X} 
\and Igor~Yu.~Lapshov\inst{\ref{in:IKI}}\orcidlink{0009-0000-4769-452X}  
\and Alexander~A.~Lutovinov\inst{\ref{in:IKI},\ref{in:RAS}}\orcidlink{0000-0002-6255-9972}   
\and Alexey~Yu.~Tkachenko\inst{\ref{in:IKI}}\orcidlink{0000-0002-7486-1730}
\and Hua Feng\inst{\ref{in:IHEP}}\orcidlink{0000-0001-7584-6236}
\and Shuang-Nan Zhang\inst{\ref{in:IHEP}}\orcidlink{0000-0001-5586-1017}
\and Xiao-Bo Li\inst{\ref{in:IHEP}}\orcidlink{0000-0003-4585-589X}
\and Shu Zhang\inst{\ref{in:IHEP}} 
\and Juri Poutanen\inst{\ref{in:UTU}}\orcidlink{0000-0002-0983-0049}}
\institute{
Department of Physics and Astronomy, 20014 University of Turku,  Finland 
\label{in:UTU} 
\and
State Key Laboratory of Particle Astrophysics, Institute of High Energy Physics, Chinese Academy of Sciences, Beijing 100049, China \label{in:IHEP}
\and Space Research Institute, Russian Academy of Sciences, Profsoyuznaya str. 84/32, 117997 Moscow, Russia 
\label{in:IKI}
\and School of Physics, Nankai University,  Tianjin 300071, China
\label{in:NU}
\and
Institut f\"ur Astronomie und Astrophysik,
Kepler Center for Astro and Particle Physics,
Eberhard Karls Universit\"at T\"ubingen,
Sand~1, 72076 T\"ubingen, Germany
\label{in:IAAT}
\and Russian Academy of Sciences, Leninsky prosp. 14, 119991 Moscow, Russia 
\label{in:RAS}
\and 
University of Chinese Academy of Sciences, Chinese Academy of Sciences, Beijing 100049, China
\label{in:UCAS}
\and 
School of Physics and Astronomy, Sun Yat-Sen University, Zhuhai 519082, P.R. China
\label{in:SYS}
\and
Mullard Space Science Laboratory, University College London, Holmbury St. Mary, Surrey RH5 6NT, UK
\label{in:UCL}
\and
Astrophysics, Department of Physics, University of Oxford, Denys Wilkinson Building, Keble Road, Oxford OX1 3RH, UK
\label{in:OXF}
}
\date{Received XXX}

\abstract
{Quasi-periodic oscillations (QPOs) are commonly observed in accreting X-ray pulsars (XRPs), but the physical mechanisms governing their frequency evolution remain poorly understood.}
{We investigate the evolution of the QPO during the 2026 outburst of the Be/X-ray pulsar \src, focusing on its frequency--luminosity relation during the rising and declining phases of the outburst.}
{We analyzed X-ray timing and spectral observations obtained with \srg/ART-XC, \nustar, and \hxmt. We measured QPO centroid frequencies and amplitudes, compared their evolution during the rising and declining phases of the outburst, and examined the associated changes in the X-ray spectral shape.}
{We detected a prominent QPO with a centroid frequency ranging from approximately 37 to 114 mHz. Its frequency–luminosity relation follows two distinct branches, with QPO frequencies approximately 35\% higher during the rise than during the decline at the same luminosity. The frequency evolution along both branches is consistent with Keplerian motion near the inner disk edge. The X-ray spectral shape also follows different evolutionary tracks during the rising and declining phases.}
{We report the first evidence for QPO frequency hysteresis in an XRP. The observed separation between the rising and declining branches suggests that the characteristic timescale of the QPO depends not only on the instantaneous luminosity but also on the accretion history. We discuss possible implications for the disk--magnetosphere interaction.}

\keywords{accretion, accretion disks -- magnetic fields -- pulsars: individual: \src\ -- stars: neutron -- X-rays: binaries}

\maketitle
\nolinenumbers
\section{Introduction}

Quasi-periodic oscillations (QPOs) are observed in a variety of accreting systems, including black-hole X-ray binaries \citep{Ingram2019,Marcel2026}, X-ray pulsars (XRPs; \citealt{Manikantan2024,MushtukovTsygankov2024}), and accreting white dwarfs \citep{Veresvarska2024}. 
In strongly magnetized XRPs, QPO frequencies span a broad range, from below 1~mHz to the Hz regime, potentially reflecting different physical mechanisms operating in the accretion flow.

At the lowest frequencies, QPOs have been detected in the ultraluminous XRPs M51~ULX-7 (0.5--0.6~mHz; \citealt{Imbrogno2024}) and M82~X-2 (3--4~mHz; \citealt{Feng2010}), as well as in the near-Eddington Be/X-ray pulsar (BeXRP) SXP31.0 (0.8~mHz; \citealt{Salganik2026SXP31}). 
One possible explanation for such low-frequency oscillations is magnetically driven warping and precession of the inner accretion disk \citep{ShirakawaLai2002}. 
This mechanism has been explored in the context of ultraluminous XRPs accreting at super-Eddington rates \citep{Veresvarska2025} and has also been considered for the 0.8~mHz QPO in SXP31.0 \citep{Salganik2026SXP31}.

At higher frequencies, QPOs have been observed in several BeXRPs, including 4U~0115+63 (27--46~mHz; \citealt{Dugair2013}), 1A~0535+262 (up to 94~mHz; \citealt{Kong2026}), and XTE~J1858+034 (180~mHz; \citealt{Manikantan2024}). 
An even higher-frequency QPO at 1.27~Hz was reported in SXP31.0 during its 1998 outburst \citep{Kaur2007}, although this feature was not recovered during the subsequent outburst studied by \citet{Salganik2026SXP31}. 

Two commonly invoked interpretations of QPOs in accreting XRPs are the Keplerian frequency model (KFM), which associates oscillations with orbital motion of inhomogeneities near the inner edge of the accretion disk \citep{vanderKlis1987}, and the magnetospheric beat frequency model (BFM), in which the observed frequency corresponds to the difference between the local Keplerian frequency and the neutron-star (NS) spin frequency \citep{Alpar1985}. 
In both models, the QPO frequency is related to the disk truncation radius, which depends on the mass accretion rate and the NS magnetic field. For a dipolar magnetic field and a constant disk--magnetosphere coupling factor, the KFM predicts $\nu_{\rm QPO}\propto L^{3/7}$, with approximately the same scaling expected in the BFM when the spin-frequency correction is negligible.

However, the observed frequency--luminosity behavior is not always consistent with these simple predictions. Multiple QPOs at distinct frequencies, occasionally occurring simultaneously, have been detected in 4U~0115+63, with no clear correlation between their frequencies and X-ray flux \citep{Dugair2013}. A similar lack of correlation has been reported in \mbox{Cen~X-3} \citep{Raichur2008}. By contrast, recent spectral--timing observations of 1A~0535+262 revealed a QPO frequency--luminosity relation, $\nu_{\rm QPO}\propto L^{0.62\pm0.01}$, steeper than predicted by the standard Keplerian frequency model, along with QPO-phase-dependent variations in the cyclotron line \citep{Kong2026}. The latter suggests a connection between the quasi-periodic variability and the physical conditions in the line-forming region, although the oscillations themselves may originate elsewhere in the accretion flow.

The BeXRP \src exhibited QPOs at frequencies of approximately 0.07--0.09~Hz during its 2009 outburst \citep{Devasia2011,Maitra2012}. The QPO frequency decreased from approximately 0.09 to 0.07~Hz as the outburst declined, suggesting a connection between the oscillations and the evolving accretion conditions \citep{Devasia2011}. Broad-band X-ray spectroscopy of the same outburst revealed a cyclotron resonant scattering feature (CRSF) near 55~keV \citep{Doroshenko2010,Suchy2011,Maitra2012}. The detection of both QPOs and a CRSF makes \src a particularly interesting system for investigating the relationship between variability in the inner accretion flow and the physical conditions in the X-ray-emitting region.

The source was first detected by \textit{Ariel-5} during a month-long outburst in 1974  and subsequently identified as an XRP with a spin period of $405.3\pm0.6$~s \citep{Ives1975,Fabian1975}. 
Its optical counterpart, classified as an O9.5\,IV--Ve star, established its Be/X-ray binary nature \citep{Chevalier1975,JanotPacheco1981}. 
Earlier distance estimates of 3--7~kpc \citep{Coe1985,JanotPacheco1981} have been revised to $2.9\pm0.1$~kpc based on Gaia DR3 \citep{GaiaDR3,Nuchvanichakul2025}. 
The orbital solution derived by \citet{Staubert2011} yielded $P_{\rm orb}=24.0\pm0.4$~d and constrained the remaining orbital parameters. 
Major outbursts were also observed in 1992 and 2009, each lasting approximately one month \citep{Coe1994,Mangano2009a,Mangano2009b,Doroshenko2010}.

On 2026 January 17, \src entered a new major X-ray outburst \citep{ATel17608,GCN43442,GCN43435,ATel17613,Molkov2026}. 
Optical spectroscopy with the Southern African Large Telescope (SALT) revealed prominent Balmer emission lines, indicating an active circumstellar disk \citep{ATel17620}. 
Three \nustar observations covered the bright, intermediate, and low-luminosity stages of the outburst \citep{Salganik2026}. 
A QPO at 0.106~Hz was reported during the bright phase \citep{Chhotaray2026}. 
The luminosity-dependent spectral evolution was also investigated by \citet{Salganik2026}, who reported the emergence of a two-hump continuum as the luminosity decreased. 
The recurrence of QPOs across different outbursts, together with the broad luminosity coverage and pronounced spectral evolution observed in 2026, provides an opportunity to investigate how quasi-periodic variability is related to changes in the inner accretion flow and the X-ray-emitting region.

In this paper, we study the 2026 outburst of \src\ and trace the evolution of its QPO frequency and spectral properties, focusing on their luminosity dependence. 
The observations and data reduction are presented in Sect.~\ref{sec:observations}, the results in Sect.~\ref{sec:results}, their interpretation in Sect.~\ref{sec:discussion}, and the conclusions in Sect.~\ref{sec:conclusions}.

\section{Observations and data reduction}
\label{sec:observations}

\subsection{\nustar}

The \nustar\ observatory operates in the 3--79~keV energy range and is equipped with two nearly identical, co-aligned focal plane modules, FPMA and FPMB \citep{Harrison2013}. 
During the 2026 outburst, \src\ was observed on three occasions: 2026 January~24 (MJD 61064.1--61064.4; ObsID 91201305002), February~21 (MJD 61092.4--61093.0; ObsID 91201310002), and March~22 (MJD 61121.8--61123.3; ObsID 91202318002). 

The data were reduced following the standard \nustar\ analysis procedures,\footnote{\url{https://heasarc.gsfc.nasa.gov/docs/nustar/analysis/nustar_swguide.pdf}} using \textsc{HEASoft} v6.36 and CALDB version 20260407. 
To suppress enhanced particle background during passages through the South Atlantic Anomaly (SAA), we applied the recommended \texttt{nupipeline} filtering settings: \texttt{saacalc=1}, \texttt{saamode=OPTIMIZED}, and \texttt{tentacle=yes}. 
For the brightest observation (NuObs1), an additional filtering criterion, \texttt{(STATUS==b0000xxx00xxxx000)\&(SHIELD==0)}, was imposed to minimize possible background contamination at high count rates. 
After filtering, the effective exposure times were 13.2, 27.6, and 58.7~ks for NuObs1, NuObs2, and NuObs3, respectively.

Source photons were collected from circular regions centered on the source, with radii of $60\arcsec$ for NuObs1 and $50\arcsec$ for both NuObs2 and NuObs3. 
The background was estimated from a source-free circular region of radius $120\arcsec$. 
Both spectra and light curves were extracted using the \texttt{nuproducts} task within the \texttt{nustardas} pipeline. 
To improve the counting statistics, the background-subtracted light curves obtained independently with FPMA and FPMB were subsequently combined using the \texttt{lcmath} tool.

\subsection{\textit{SRG}/ART-XC}

The \srg\ observatory, launched in 2019 from the Baikonur Cosmodrome into an orbit around the Sun--Earth L2 point, is equipped with two X-ray telescopes, including the Mikhail Pavlinsky ART-XC instrument \citep{2021A&A...656A.132S}. 
ART-XC comprises seven co-aligned telescope modules, each combining an X-ray mirror assembly with a double-sided CdTe strip detector \citep{2021A&A...650A..42P}. 
Although the telescope is optimized for observations in the 4--30~keV band, its sensitivity extends to energies of approximately 120~keV. 
For the spectral flux measurements presented in this work, we adopted the 4--35~keV energy range to maximize coverage of the high-energy hump while maintaining sufficient instrumental sensitivity.

During the 2026 outburst of \src, \srg/ART-XC performed 22 observations between January~19 and April~1 (ObsIDs 12610180001--12610180022). 
The observations were reduced using the \textsc{artproducts} software package (v1.0) with the CALDB v20230228 calibration files. 
Source events were extracted from a circular region of radius $2\farcm5$ centered on the target, with energy filtering applied where necessary. 
The background was estimated using a concentric annulus with inner and outer radii of $4\arcmin$ and $7\arcmin$, respectively. 

\begin{table*}[t]
\centering
\caption{QPO centroid frequencies and X-ray luminosities of \src\ during the 2026 outburst.}
\label{tab:qpo_properties}
\begin{tabular}{llcccccc}
\hline\hline
ObsID & Short name & MJD interval & Phase
& $L$ ($10^{37}$ erg s$^{-1}$)
& $F_{\rm high}/F_{\rm low}$
& $\nu_{\rm QPO}$ (mHz) & $r_{\rm QPO}$ (\%) \\
\hline
\multicolumn{8}{c}{\srg/ART-XC}\\
12610180001 & A01 & 61059.5--61062.3 & Rise
& $0.981\pm0.003$ & $2.26\pm0.05$ &$84.6\pm0.5$ & $4.5\pm0.1$\\

12610180002 & A02 & 61064.5--61065.4 & Rise
& $2.364\pm0.006$ &$3.28\pm0.10$& $112.5\pm0.5$ & $10.0\pm0.2$ \\

12610180003 & A03 & 61066.5--61068.6 & Transition
& $2.465\pm0.006$ &$3.55^{+0.11}_{-0.10}$& $106.6\pm0.3$ & $10.5\pm0.1$ \\

 & A03.1 & 61066.5--61067.0 & Transition
 & $2.477\pm0.007$ &$3.63\pm0.13$& $113.8\pm0.6$ & $9.8\pm0.3$ \\

 & A03.2 & 61067.0--61067.6 & Transition
 & $2.517\pm0.007$ &$3.50\pm0.12$& $111.6\pm0.7$ & $9.7\pm0.3$ \\

 & A03.3 & 61067.6--61068.1 & Transition
 & $2.476\pm0.007$ &$3.52\pm0.12$& $106.2^{+0.5}_{-0.6}$ & $10.4\pm0.3$ \\

 & A03.4 & 61068.1--61068.6 & Transition
 & $2.410\pm0.007$ &$3.58_{-0.12}^{+0.13}$& $99.4\pm0.5$ & $10.6\pm0.3$ \\

12610180004 & A04 & 61069.6--61070.6 & Decline
& $2.172\pm0.005$ &$3.78_{-0.12}^{+0.13}$& $85.9\pm0.4$ & $10.9\pm0.2$ \\

12610180005 & A05 & 61072.8--61073.2 & Decline
& $1.691\pm0.005$ &$3.75_{-0.13}^{+0.14}$& $74.2\pm0.6$ & $11.4\pm0.3$ \\

12610180006 & A06 & 61074.5--61075.0 & Decline
& $1.344\pm0.004$ & $3.37\pm0.12$& $66.9\pm0.4$ & $10.5\pm0.3$\\

12610180007 & A07 & 61075.9--61076.4 & Decline
& $1.127\pm0.004$ &$3.11_{-0.10}^{+0.11}$& $59.3\pm0.6$ & $10.9\pm0.3$ \\

12610180008 & A08 & 61077.7--61078.2 & Decline
& $0.898\pm0.003$ &$2.88\pm0.09$& $55.9\pm0.5$ & $9.7\pm0.3$ \\

12610180009 & A09 & 61079.1--61079.6 & Decline
& $0.736\pm0.003$ &$2.71_{-0.08}^{+0.09}$& $51.8\pm0.5$ & $8.1\pm0.3$ \\

12610180010 & A10 & 61081.5--61082.0 & Decline
& $0.489\pm0.002$ &$2.49_{-0.07}^{+0.08}$& $43.8^{+1.1}_{-0.9}$ & $4.2_{-0.5}^{+0.3}$\\

12610180011 & A11 & 61082.9--61083.9 & Decline
& $0.318\pm0.001$ &$1.95\pm0.05$& $37.1^{+0.6}_{-0.5}$ & $3.9_{-0.1}^{+0.2}$ \\

\hline
\multicolumn{8}{c}{\nustar}\\

91201305002 & NuObs1 & 61064.1--61064.4 & Rise
& $1.985^{+0.002}_{-0.001}$ & $2.76\pm0.25$& $106.5\pm1.2$ & $7.0\pm0.5$\\

\hline
\multicolumn{8}{c}{\hxmt}\\

P0804763005 & 3005 & 61064.4--61064.8 & Rise
& $2.07\pm0.02$ & --
& $117.7^{+4.5}_{-3.1}$
& $2.9\pm0.5$ \\

P0804763006 & 3006 & 61065.4--61066.0 & Rise
& $2.44\pm0.02$ & --
& $117.0^{+1.8}_{-1.6}$
& $3.4^{+0.6}_{-0.5}$ \\

P0804763007 & 3007 & 61066.6--61067.1 & Transition
& $2.31\pm0.02$ & --
& $119.1^{+2.7}_{-2.6}$
& $12.9^{+1.3}_{-1.5}$ \\

P0804763008 & 3008 & 61067.5--61067.9 & Transition
& $2.62\pm0.02$ & --
& $106.3^{+2.6}_{-2.8}$
& $7.7\pm1.1$ \\

P0804557002 & 7002 & 61068.7--61069.1 & Decline
& $2.60\pm0.02$ & --
& $92.9^{+1.5}_{-2.1}$
& $8.1^{+1.3}_{-1.1}$ \\

P0804557005 & 7005 & 61069.6--61070.0 & Decline
& $2.56\pm0.03$ & --
& $86.8^{+1.6}_{-1.7}$
& $9.8\pm0.7$ \\

P0804557006 & 7006 & 61070.8--61071.2 & Decline
& $2.03\pm0.02$ & --
& $78.1^{+1.2}_{-1.3}$
& $10.1\pm0.8$ \\

P0804557007 & 7007 & 61071.8--61072.2 & Decline
& $2.21\pm0.02$ & --
& $72.0\pm2.7$
& $9.3^{+1.8}_{-1.9}$ \\

P0804557008 & 7008 & 61073.1--61073.5 & Decline
& $1.54\pm0.02$ & --
& $73.7\pm1.6$
& $11.2\pm1.2$ \\

P0804557009 & 7009 & 61074.1--61074.5 & Decline
& $1.77\pm0.02$ & --
& $66.0\pm0.6$
& $6.7\pm0.9$ \\

\hline

\hline
\end{tabular}
\tablefoot{
The MJD intervals indicate the start and end times of the full observations.
Luminosities are given in the 4--35~keV energy band.
The four A03 subsegments are listed in chronological order.
Their luminosities were determined from independent
time-resolved spectral fits. The flux ratio $F_{\rm high}/F_{\rm low}$ is calculated
from the intrinsic 3--79~keV fluxes of the two
Comptonization components following the spectral
analysis of \citet{Salganik2026}.
}
\end{table*} 
\subsection{\hxmt}
\label{sec:hxmt}
The Hard X-ray Modulation Telescope (\hxmt), launched on 2017 June 15, is China's first space-based X-ray astronomy observatory \citep{Zhang2020}. The satellite hosts three instruments covering complementary energy bands: the High Energy (HE) telescope operating at 20--250~keV \citep{Liu2020}, the Medium Energy (ME) telescope covering 5--30~keV \citep{Cao2020}, and the Low Energy (LE) telescope sensitive to 1--15~keV \citep{Chen2020}. Each instrument comprises multiple detector units with different fields of view and pointing orientations. The total effective areas are approximately 5000, 952, and 384~cm$^2$ for HE, ME, and LE, respectively. During the 2026 outburst of \src, \hxmt\ carried out 26  pointed observations between January~21 and February~24.
\begin{figure}[t]
    \centering
    \includegraphics[width=\columnwidth]{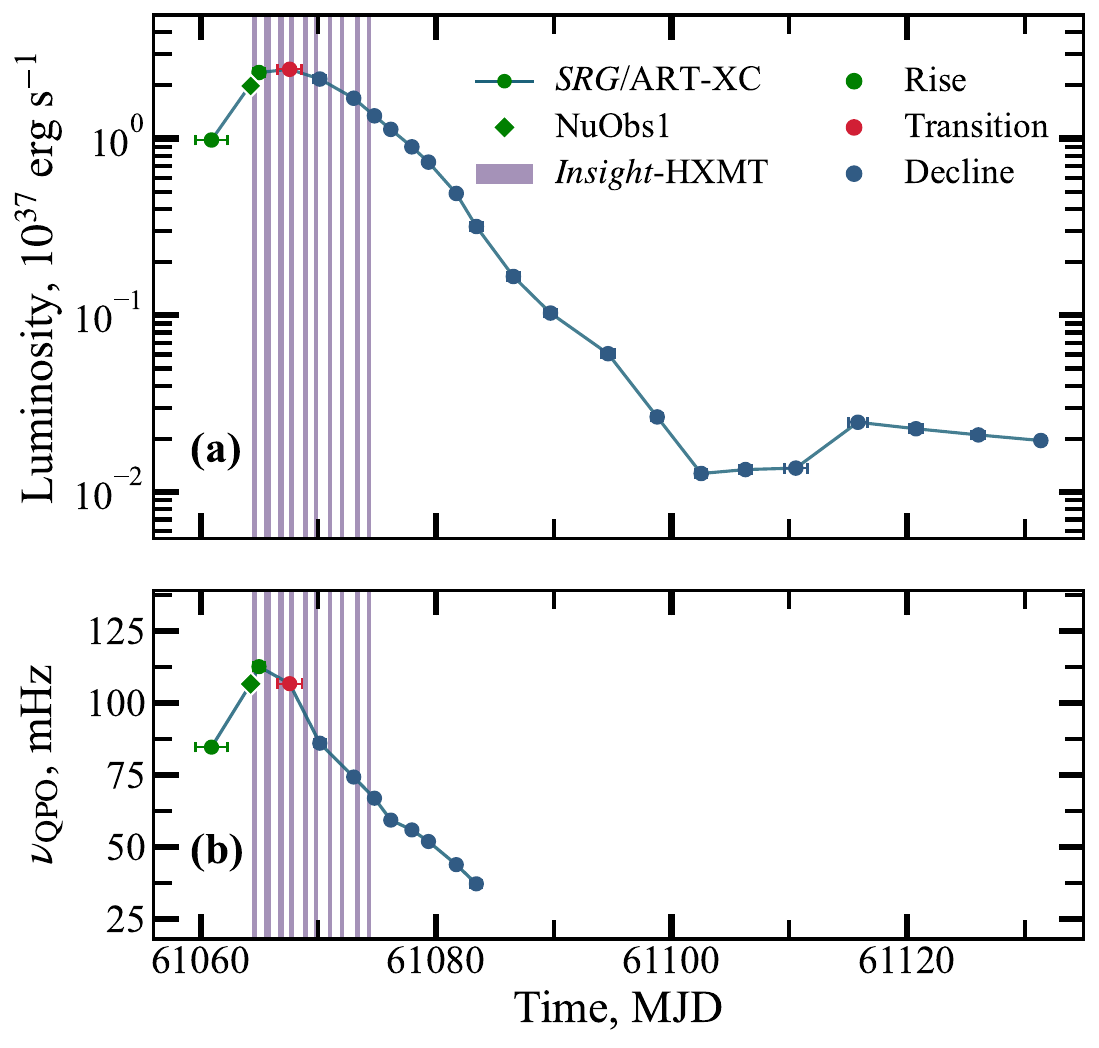}
    \caption{
Long-term evolution of \src\ during the 2026 outburst.
Panel (a): 4--35~keV X-ray luminosity as a function of time.
Panel (b): QPO centroid frequency as a function of time.
Green, red, and blue symbols correspond to the rising phase,
transitional observation A03, and declining phase,
respectively.
The purple shaded regions indicate the time intervals of
the selected \hxmt\ observations.
    }
    \label{fig:longterm_evolution}
\end{figure}

The data were processed with the \hxmt\ Data Analysis Software (\textsc{HXMTDAS} v2.07)\footnote{\url{http://hxmtweb.ihep.ac.cn/software.jhtml}} and CALDB version 2.08.\footnote{\url{http://hxmtweb.ihep.ac.cn/SoftDoc/847.jhtml}} Good time intervals (GTIs) were determined using the \texttt{he/me/legtigen} tasks by requiring an Earth elevation angle greater than $10\degr$, a time separation from passages through the South Atlantic Anomaly (SAA) exceeding 300~s, a geomagnetic cutoff rigidity above 8~GV, and angular distances from the Moon and Sun greater than $1\degr$ and $70\degr$, respectively. The event lists were then filtered using the \texttt{he/me/lescreen} tasks, and photon arrival times were corrected to the Solar System barycenter with \texttt{hxbary}. Light curves and spectra were generated using \texttt{he/me/lelcgen} and \texttt{he/me/lespecgen}, respectively, while the corresponding background estimates were obtained using the \texttt{he/me/lebkgmap} tasks.

The \hxmt\ observations are susceptible to source
contamination within the instrument field of view,
particularly from Cen~X-3. This contamination can
introduce systematic biases in the measured X-ray
luminosities and may also affect the QPO centroid
frequencies through additional variability from
the contaminating source. We therefore exclude
the \hxmt\ measurements from the quantitative
frequency--luminosity analysis and use them
for qualitative comparison only.

\begin{figure}
\centering
\includegraphics[width=1.0\columnwidth]{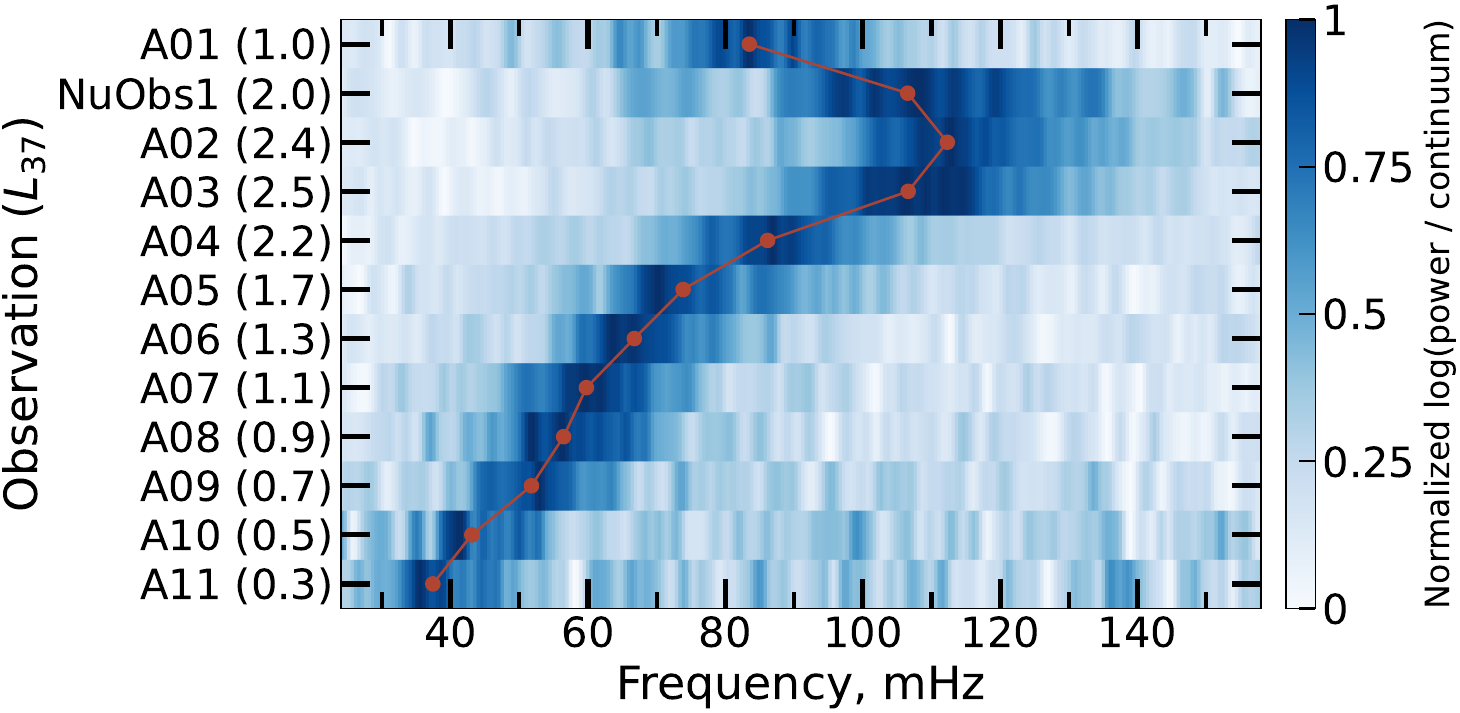}
\caption{Evolution of the PDS of \src\ during the 2026 outburst, as observed with ART-XC and \nustar.
Each horizontal strip represents an individual observation, with the corresponding 4--35~keV luminosity indicated in units of $10^{37}$~\lum.
The power spectra are divided by their best-fit broad-band noise continua and Gaussian-smoothed along the frequency axis for visualization purposes only.
The color scale represents the logarithm of the power-to-continuum ratio, normalized independently for each observation.
The red circles indicate the measured QPO centroid frequencies.}
    \label{fig:qpo_pds_evolution}
\end{figure}

\begin{figure*}
\centering
\includegraphics[width=0.8\textwidth]{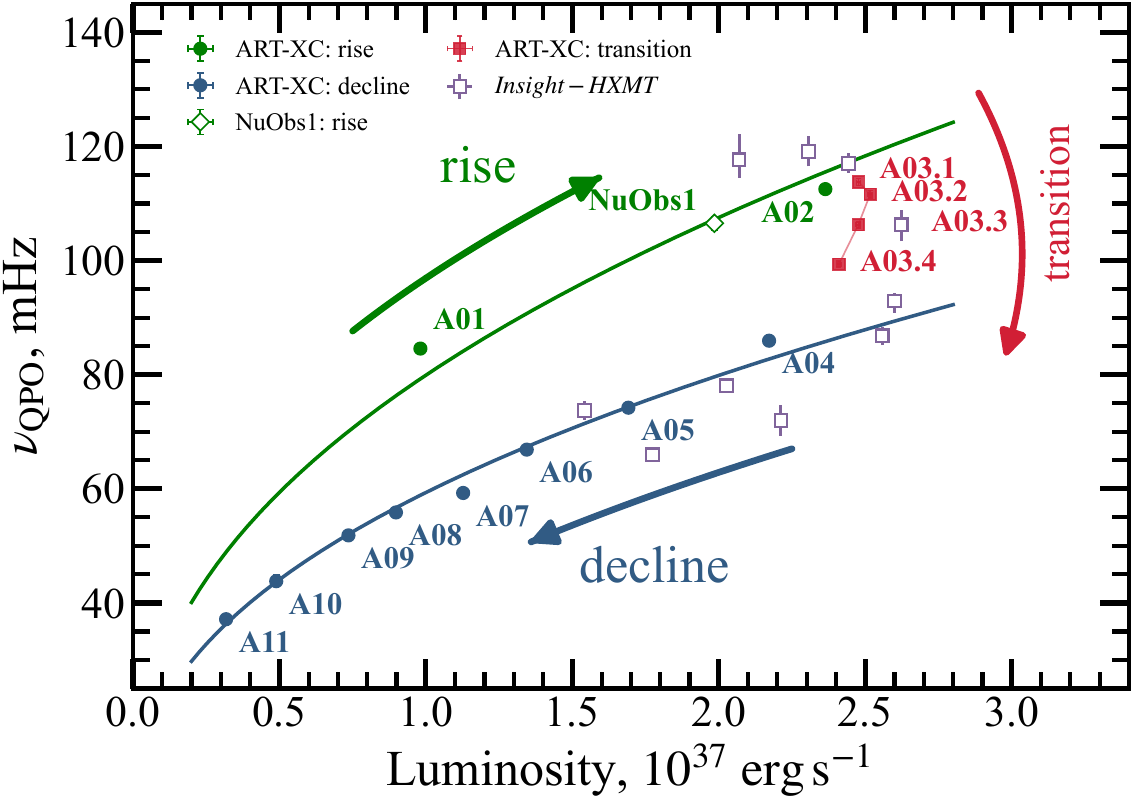}
\caption{QPO centroid frequency as a function of the 4--35~keV luminosity during the 2026 outburst of \src.
All ART-XC and \nustar observations are represented by their observation-averaged measurements, except for ART-XC observation A03, which is divided into four consecutive time intervals (A03.1--A03.4).
The \hxmt measurements are overplotted for comparison only.
Solid lines represent the best-fit relations $\nu_{\rm QPO}=\mathcal{C}_{\rm b}L_{37}^{3/7}$.}
\label{fig:qpo_peak_drift}
\end{figure*}

\subsection{Spectral data approximation}
\label{sec:spec_approx} 

The spectral analysis followed the procedure described in detail by \citet{Salganik2026}.
The observation-averaged ART-XC and \nustar\ spectral measurements were adopted from that work.
All spectra were grouped using the optimal binning method of \citet{Kaastra2016}, implemented with the \texttt{optmin} option, requiring at least 1, 25, and 25 counts per bin for \nustar, ART-XC, and \hxmt, respectively.
The \nustar\ spectra were analyzed using the W-statistic \citep{Wachter1979}, whereas the ART-XC and \hxmt\ spectra were fitted using the $\chi^2$ statistic.
Systematic uncertainties of 1.5\% and 1\% were included for ART-XC and \hxmt, respectively.

For the four subsegments of ART-XC observation A03, we extracted and fitted independent spectra in the 5--35~keV energy range using \textsc{xspec} \citep{Arnaud1996} with the model \texttt{tbabs$\times$(comptt+comptt+gaussian)}.
Following \citet{Salganik2026}, the seed-photon temperature $kT_0$ and the electron temperature of the low-energy Comptonization component, $kT_{\rm e,low}$, were fixed to values interpolated between the three \nustar\ measurements in $\log L_{4-35}$. 
The optical depth of the high-energy component was fixed at $\tau_{\rm high}=200$.
The unabsorbed 4--35~keV luminosities were determined independently for each subsegment from its best-fit spectral model, without count-rate scaling.

We also performed spectral fits of the \hxmt\ observations using the two-component Comptonization  prescription of \citet{Salganik2026}.
The seed-photon temperature was fixed using the same luminosity-dependent interpolation procedure.
The resulting models were used to estimate the unabsorbed 4--35~keV luminosities.

Following \citet{Salganik2026}, the intrinsic 3--79~keV fluxes of the two Comptonization components were evaluated separately using \texttt{cflux}.
We adopt the ratio $F_{\rm high}/F_{\rm low}$,
the inverse of that presented in the earlier work.
For ART-XC, these fluxes partly rely on model
extrapolation beyond the fitted energy range.
The component flux ratio was not determined for
\hxmt\ because source contamination prevents a
reliable decomposition of the continuum.
All luminosities were calculated assuming
a distance of 2.9~kpc. Unless otherwise stated, all uncertainties quoted in this paper correspond to the $1\sigma$ confidence level.

\section{Results}
\label{sec:results}
\subsection{QPO analysis}
To investigate the aperiodic X-ray variability of \src\ during the 2026 outburst, we constructed power density spectra (PDS) and cospectra from observations with \srg/ART-XC, \hxmt, and \nustar. The spectra were modeled using \textsc{xspec} with a power-law continuum and two Lorentzian components (\texttt{powerlaw+lorentz+lorentz}), describing the broad-band noise and the QPO feature. 

For \srg/ART-XC and \hxmt, we constructed PDS using the \texttt{powspec} task from \textsc{HEASoft}. The ART-XC PDS were calculated from barycenter-corrected events in the 4--12~keV energy band with a time resolution of 0.125~s, while the \hxmt\ PDS were obtained from 0.2-s-binned light curves in the 7--35~keV ME band, except for observation P0804763007, in which the QPO was detected only in the 2--10~keV LE band. For \hxmt, we used 2048 time bins per Fourier interval. The individual Fourier spectra were averaged for each observation without frequency rebinning. The PDS were normalized in units of fractional rms$^2$ Hz$^{-1}$, with the expected white-noise level subtracted (normalization = $-2$ in \texttt{powspec}), and subsequently converted into an \textsc{xspec}-compatible format using the \texttt{flx2xsp} utility.

The ART-XC PDS were fitted in the 0.01--0.60~Hz frequency range, excluding Fourier bins corresponding to the coherent pulsation frequency and its harmonics. For \hxmt, the fitting range was restricted to 0.01--0.15~Hz to avoid contamination by coherent pulsations of the nearby XRP Cen~X-3 at approximately 0.2~Hz (Sect.~\ref{sec:hxmt}). The fits were performed using the Whittle likelihood \citep[e.g.,][]{Whittle1953,Whittle1957}, accounting for the number of averaged Fourier spectra contributing to each frequency bin. The QPO fractional rms amplitude, $r_{\rm QPO}$, was calculated as the square root of the normalization of the corresponding Lorentzian component.\footnote{\url{https://heasarc.gsfc.nasa.gov/docs/software/xspec/manual/node229.html}} 

For NuObs1, to mitigate instrumental dead-time effects we calculated the cospectrum between FPMA and FPMB in the 4--12~keV energy band using \textsc{Stingray} \citep{Bachetti2015,Huppenkothen2019}. 
The cospectra were computed with a time resolution of 0.125~s and averaged over 33 uninterrupted 512-s segments within the common good-time intervals. 
The resulting cospectrum was fitted in the 0.02--0.30~Hz frequency range using the same continuum and Lorentzian model. 
The QPO fractional rms amplitude was derived from the integrated Lorentzian power.

\subsection{QPO luminosity dependence}
\label{sec:qpo_luminosity}

A prominent QPO was detected in 11 \srg/ART-XC observations (A01--A11), one \nustar observation (NuObs1), and ten \hxmt\ observations. 
The measured QPO properties and corresponding luminosities are summarized in Table~\ref{tab:qpo_properties}.

Figure~\ref{fig:longterm_evolution} shows the temporal evolution of the X-ray luminosity and QPO centroid frequency during the 2026 outburst of \src. 
During the rise, the QPO frequency increased from $84.6\pm0.5$~mHz in A01 to $112.5\pm0.5$~mHz in A02. 
The \nustar\ measurement of $106.5\pm1.2$~mHz is consistent with this trend and with the previously reported QPO at 106~mHz \citep{Chhotaray2026}. 
During the decline, the frequency decreased from $85.9\pm0.4$~mHz in A04 to $37.1^{+0.6}_{-0.5}$~mHz in A11. 
The corresponding evolution of the PDS is illustrated in Fig.~\ref{fig:qpo_pds_evolution}.

Despite this overall correlation with luminosity, the QPO frequency followed distinct tracks during the rising and declining phases. 
At comparable luminosities, it was systematically lower during the decline, as illustrated by the A01--A08 and A02--A04 observation pairs (Fig.~\ref{fig:qpo_peak_drift}). 
This separation provides clear evidence for hysteresis in the QPO frequency--luminosity relation.

To quantify hysteresis, we fitted 11 independent measurements from ART-XC (A01--A02 and A04--A11) and \nustar\ (NuObs1) with a power law, $\nu_{\rm QPO}=\mathcal{C}_{\rm b}L_{37}^{p}$, where $L_{37}$ is the 4--35~keV luminosity in units of $10^{37}$~erg~s$^{-1}$ and $\mathcal{C}_{\rm b}$ is branch-dependent normalization in mHz. 
A fit with a common power-law index and separate normalizations yielded $p=0.41\pm0.03$, $\mathcal{C}_{\rm rise}=80.9\pm1.7$~mHz, and $\mathcal{C}_{\rm decline}=59.6\pm0.9$~mHz. 
The fitted index is consistent with the KFM prediction of $p=3/7$. 
Fixing $p=3/7$, we obtained $\mathcal{C}_{\rm rise}=79.9\pm2.0$~mHz and $\mathcal{C}_{\rm decline}=59.3\pm0.6$~mHz, corresponding to QPO frequencies approximately 35\% higher during the rise than during the decline at the same luminosity (Fig.~\ref{fig:qpo_peak_drift}).

\begin{figure} 
\centering
\includegraphics[width=\columnwidth]{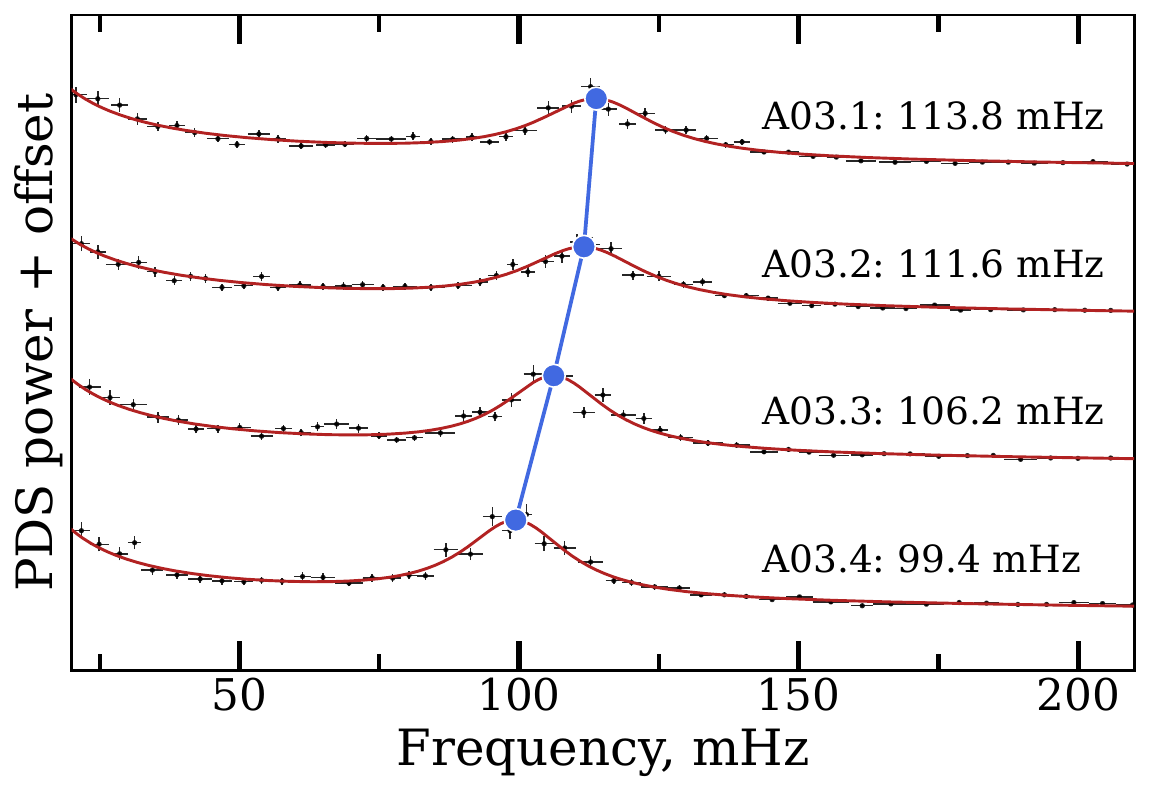}
\caption{Evolution of the PDS during ART-XC observation A03 in the 4--12~keV energy band.
The four consecutive intervals are arranged chronologically from top to bottom, with vertical offsets for clarity.
Gray points show the observed PDS, rebinned for visualization only, while red curves represent the best-fit models.
Blue circles mark the fitted QPO centroid frequencies.
The corresponding frequencies are indicated next to each PDS.}
    \label{fig:a03_pds_stacked}
\end{figure}

To trace the transition between the branches near the outburst maximum, we divided ART-XC observation A03 into four consecutive time intervals with approximately equal usable exposure for the PDS analysis.
We fitted their PDS and independently extracted spectra to measure the 4--35~keV luminosities. 
The QPO frequency decreased from $113.8\pm0.6$ to $99.4\pm0.5$~mHz (Fig.~\ref{fig:a03_pds_stacked}), while the luminosity initially increased to $2.5\times10^{37}$~\lum before decreasing to $2.4\times10^{37}$~\lum. 
Thus, the frequency declined by 13\% (by 14.4~mHz) while the luminosity decreased by only 3\%; the Keplerian scaling $\nu_{\rm QPO}\propto L^{3/7}$ would predict a decrease of merely 1\% (1.3~mHz). 
The four A03 intervals therefore trace a rapid transition between the branches (Fig.~\ref{fig:qpo_peak_drift}) and were excluded from the branch fits. 

The ten \hxmt\ QPO measurements are broadly consistent with the two branches. 
Near the outburst maximum, \hxmt\ independently detected a decrease from $119.1$ to $92.9$~mHz over approximately two days, qualitatively consistent with the rapid transition seen in A03. 
The \hxmt\ measurements are shown for comparison only and were excluded from the fits because of source contamination (Sect.~\ref{sec:hxmt}).

\subsection{Spectral--timing evolution}
\label{sec:spectral_timing}

To investigate whether the QPO frequency hysteresis is accompanied by spectral changes, we examined the luminosity dependence of the continuum flux ratio $F_{\rm high}/F_{\rm low}$ (Sect.~\ref{sec:spec_approx}). The observation-averaged component flux ratios can be obtained from \citet{Salganik2026}, where the inverse ratio was reported. The values used here, including those derived for the four intervals of A03, are listed in Table~\ref{tab:qpo_properties}. 

\begin{figure} 
\centering
\includegraphics[width=\columnwidth]{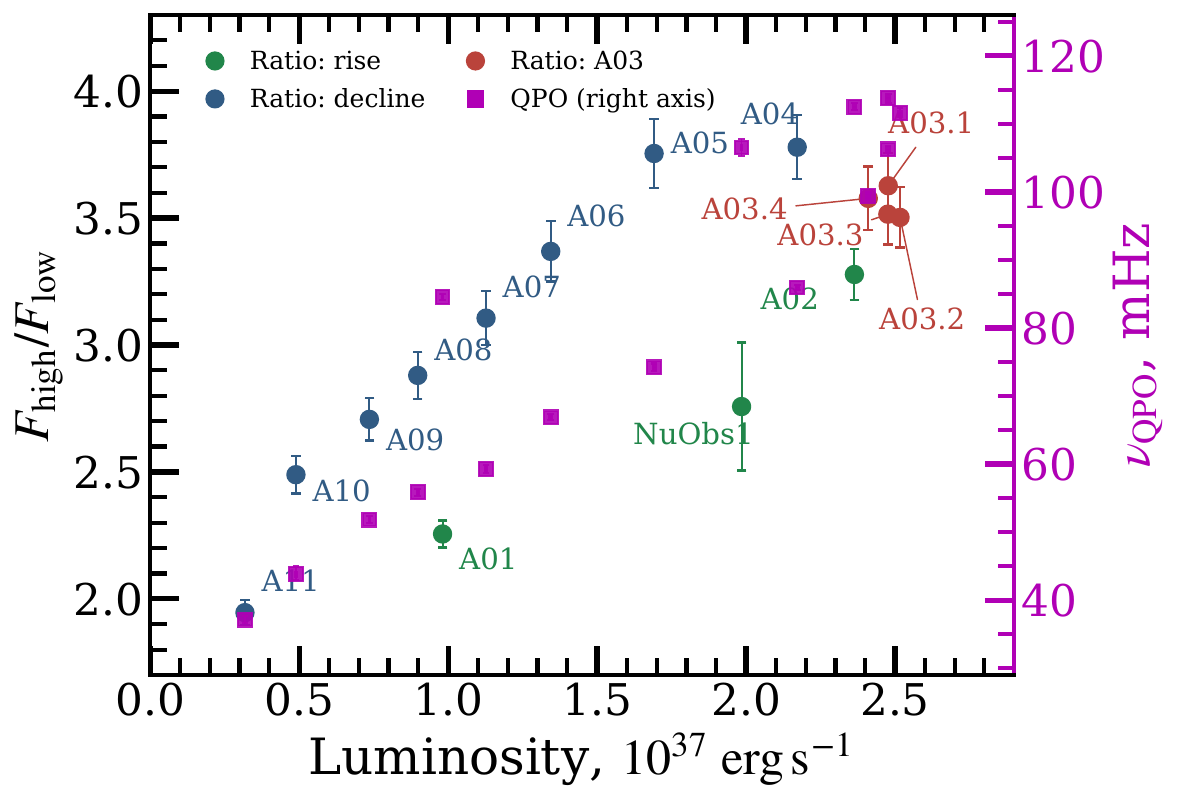}
\caption{Luminosity dependence of the continuum flux ratio
$F_{\rm high}/F_{\rm low}$ (green, blue, and red circles, left axis)
and QPO centroid frequency (magenta circles, right axis).
The ratio represents the unabsorbed 3--79~keV fluxes of the
high- and low-energy Comptonization components of the spectral
model \texttt{tbabs}$\times$(\texttt{comptt}+\texttt{comptt}+
\texttt{gaussian}).}
\label{fig:qpo_spectral_hysteresis}
\end{figure}

Figure~\ref{fig:qpo_spectral_hysteresis} shows that the continuum flux ratio follows distinct tracks during the rising and declining phases, being systematically higher during the decline at comparable luminosities, in contrast to the QPO frequency which is lower at the decline. 
For example, the ratio is $2.26\pm0.05$ in A01 while it is $2.88\pm0.09$ in A08, and $3.28\pm0.10$ in A02 while it is  $3.78^{+0.13}_{-0.12}$ in A04. 
During the transitional observation A03, the ratio remains consistent with a constant value within the uncertainties, despite the pronounced decrease in QPO frequency. 
Thus, both the spectral and timing properties exhibit hysteresis over the same luminosity range.

\subsection{Effect of the binary orbital motion}
\label{sec:qpo_orbital_uncertainty}

To assess whether orbital Doppler modulation could explain the QPO hysteresis, we estimated its maximum effect. 
For the circular orbit of \citet{Staubert2011} ($P_{\rm orb}=24$~d, $a_{\rm X}\sin i/c=54.9$~s), the maximum frequency shift at 114~mHz is only $\sim0.02$~mHz. 
Even an extreme hypothetical orbit ($P_{\rm orb}=10$~d, $a_{\rm X}\sin i/c=100$~s, $e=0.8$) would produce a difference of at most $\sim0.5$~mHz between orbital phases, far below the observed 14.4~mHz decrease during A03 (Sect.~\ref{sec:qpo_luminosity}). 
Thus, orbital Doppler modulation cannot account for the observed QPO hysteresis.

We also attempted to reproduce the measured spin-period evolution using the orbital solution of \citet{Staubert2011}, but could not obtain a satisfactory fit.
A possible contribution from history-dependent accretion torques is discussed in Sect.~\ref{sec:qpo_disk}.

\section{Discussion}
\label{sec:discussion}

\subsection{Hysteresis of the QPO frequency and its possible origin}
\label{sec:hysteresis}

Other forms of hysteresis have previously been observed in several BeXRPs, most notably in hardness--intensity diagrams (HIDs), where the rising and declining phases of an outburst follow different tracks at comparable luminosities. Such behavior has been reported for 4U~0115+63 and V~0332+53 \citep{Reig2008,Baum2017}. Hysteresis has also been reported in cyclotron-line energies \citep{Cusumano2016,Doroshenko2017}.

However, hysteresis in spectral hardness does not necessarily imply similar behavior in other observables. 
During the 2004--2005 outburst of V~0332+53, \citet{Tsygankov2010} found no evidence for hysteresis in the cyclotron-line energy--luminosity relation or pulsed fraction, despite the hysteresis observed in the hardness--intensity diagram.
In contrast, during the 2015 outburst, the cyclotron-line energy itself exhibited hysteresis, being approximately 1.5~keV lower during the decline than during the rise at comparable luminosities \citep{Cusumano2016,Doroshenko2017}.
Interestingly, the cyclotron-line energy recovered to approximately its pre-outburst value in 2016, suggesting that the observed hysteresis was associated with changes in the geometry of the line-forming region rather than a permanent decay of the stellar magnetic field \citep{Doroshenko2017}.

Hysteresis has also been detected in the aperiodic timing properties of accreting XRPs. 
During the 2020 giant outburst of 1A~0535+262, the characteristic frequencies of broad-band noise components were higher during the rise than during the decline at similar luminosities \citep{Reig2022}, accompanied by hysteresis in the CRSF energy and continuum parameters \citep{Kong2021}. In contrast, EXO~2030+375 exhibited lower characteristic noise frequencies during the rise, despite showing no significant hysteresis in the HID \citep{ReigNespoli2013}.

Our results extend this phenomenology to the centroid frequency of a distinct QPO. Although luminosity-dependent QPO frequencies have previously been observed in \src\ \citep{Devasia2011,Maitra2012} and other accreting XRPs, including 1A~0535+262 \citep{Finger1996, Ma2022, Kong2026}, a systematic separation between the rising and declining QPO frequency--luminosity tracks has not previously been established. To the best of our knowledge, the present observations therefore provide the first evidence for this type of hysteresis in XRPs.

\subsection{Implications for the disk--magnetosphere interaction}
\label{sec:qpo_disk}

The observed QPO may originate from inhomogeneities or instabilities near the disk--magnetosphere boundary. In the Keplerian frequency model (KFM), the QPO frequency traces orbital motion at the inner disk edge \citep{vanderKlis1987}, predicting $\nu_{\rm QPO}\propto L^{3/7}$ for a dipolar field and constant disk--magnetosphere coupling, consistent with the observed slope.

The magnetic field inferred from the cyclotron line at $E_{\rm cyc}\simeq55$~keV \citep{Doroshenko2010,Suchy2011,Chhotaray2026,Salganik2026} also yields Keplerian frequencies broadly consistent with the observations, given uncertainties in the NS parameters and disk--magnetosphere coupling.
Within this interpretation, the approximately 35\% higher QPO frequency during the rise than during the decline at the same luminosity (Sect.~\ref{sec:qpo_luminosity}) corresponds to an approximately 20\% larger inner disk radius during the decline, assuming the same luminosity--accretion rate relation in both phases.

The BFM, which associates the QPO frequency with the difference between the Keplerian frequency at the inner disk edge and the NS spin frequency \citep{Alpar1985}, gives similar inferred radii because the spin frequency ($\simeq2.5$~mHz) is much smaller than the observed QPO frequencies. The frequency--luminosity scaling therefore cannot distinguish between the KFM and BFM. Other mechanisms, including instabilities at the magnetospheric boundary, also remain possible.

A similar dependence of the accretion geometry on the outburst phase was suggested by \citet{Doroshenko2017} for the XRP V~0332+53, which exhibited hysteresis in the accretion torque, with a lower spin-up rate during the decline than during the rise at comparable X-ray fluxes. They proposed that the viscous evolution of the accretion disk could result in a smaller magnetospheric radius during the rising phase, qualitatively consistent with the difference inferred from the QPO hysteresis in \src. 

For a dipolar magnetic field, the polar-cap radius scales as $r_{\rm cap}\propto R_{\rm m}^{-1/2}$. The inferred 20\% larger magnetospheric radius during the decline therefore implies a roughly 17\% smaller polar-cap area and a 20\% higher mass accretion rate per unit area at fixed total $\dot M$.
The continuum flux ratio $F_{\rm high}/F_{\rm low}$ also exhibits hysteresis, being higher during the decline at comparable luminosities (Fig.~\ref{fig:qpo_spectral_hysteresis}). The higher mass accretion rate per unit polar-cap area inferred above could increase the density and resonant optical depth of the accretion flow, enhancing the reprocessing of radiation near the cyclotron resonance and strengthening the high-energy component relative to the low-energy one \citep{Mushtukov2021,Salganik2026}. This provides a possible connection between the QPO and spectral hysteresis. 

The rapid transition observed near the peak of the outburst provides a possible clue to the physical origin of the hysteresis. The radial viscous timescale near the magnetospheric boundary can be of the order of days for a standard accretion disk \citep[Eq.~5.69]{Frank2002}, comparable to the timescale of the observed transition. Changes in the inner disk structure or its magnetic coupling to the NS \citep{Lai2014} could counteract the inward motion of the disk boundary expected with increasing accretion rate at fixed disk--magnetosphere coupling.

The inferred magnetospheric hysteresis affects the spin evolution of the NS \citep{Doroshenko2017}. In an illustrative application of the Ghosh--Lamb model \citep{GhoshLamb1979}, the inferred difference in the inner disk radius corresponds to an approximately 8\% change in the accretion torque at fixed mass accretion rate. Adopting $M_{\rm NS}=1.4~M_\odot$, $R_{\rm NS}=10$~km, moment of inertia $I=10^{45}$~g~cm$^2$, and an accretion luminosity $L=GM_{\rm NS}\dot M/R_{\rm NS}=10^{37}$~\lum, this gives $\Delta\dot{\nu}\sim3\times10^{-13}$~Hz~s$^{-1}$. If sustained for approximately two weeks, such a difference could produce a frequency offset comparable to the orbital Doppler semi-amplitude ($\sim4\times10^{-7}$~Hz). This may explain the difficulties encountered in modeling the spin evolution using the orbit of \citet{Staubert2011}, suggesting that history-dependent accretion torques could complicate orbital timing analyses.

\section{Conclusions}
\label{sec:conclusions}

We investigated the evolution of QPOs during the 2026 outburst of the BeXRP \src\ using observations with  \srg/ART-XC, \nustar, and \hxmt. Our main findings are summarized as follows:

\begin{itemize}

\item We detected a prominent QPO with a centroid frequency evolving between approximately 37 and 114~mHz. The frequency increased with luminosity during the rising phase and decreased during the decline, following two distinct frequency--luminosity tracks. 

\item To the best of our knowledge, this represents the first evidence of QPO frequency hysteresis in an XRP, demonstrating that the QPO frequency is not uniquely determined by the instantaneous X-ray luminosity and suggesting that the accretion flow retains a memory of its previous evolution.

\item The two branches can be described by power-law relations with a common index consistent with the canonical $3/7$ scaling expected for Keplerian motion at the magnetospheric boundary. At the same luminosity, the QPO frequency is approximately 35\% higher during the rising phase than during the decline.

\item Near the outburst maximum, the QPO frequency decreased from $113.8$ to $99.4$~mHz (13\%), while the 4--35~keV luminosity decreased by only 3\%, much less than expected from the standard Keplerian scaling. In contrast, the continuum flux ratio $F_{\rm high}/F_{\rm low}$ remained nearly constant during this transition, despite exhibiting hysteresis between the rising and declining phases. 

\end{itemize}

\begin{acknowledgements}
This research was supported by the International Space Science Institute (ISSI) in Bern, through International Team project 25-657 `Polarimetric Insights into Extreme Magnetism’ and the  Research Council of Finland Centre of Excellence in Neutron-Star Physics (grant 374064). 
AS acknowledges support from the Jenny and Antti Wihuri Foundation (grant 00240331).
SST, SVM, IYL, AAL, AYT, and JP acknowledge support from the Ministry of Science and Higher Education of RF grant 075-15-2024-647.
AAM acknowledges support from the UKRI1169 Astronomy observation and theory small award.
HX acknowledges support from the China Scholarship Council (CSC: 202406380048).

This work is partially based on observations with the Mikhail Pavlinsky ART-XC telescope, hard X-ray instrument on board the \srg observatory. The \srg observatory was created by Roskosmos in the interests of the Russian Academy of Sciences represented by its Space Research Institute (IKI) in the framework of the Russian Federal Space Program, with the participation of Germany. The ART-XC team thanks the Roscosmos State Corporation, the Russian Academy of Sciences, and Rosatom State Corporation for supporting the ART-XC telescope, as well as the JSC Lavochkin Association and partners for manufacturing and running the Navigator spacecraft and platform.  

This work also made use of data obtained with the NuSTAR mission, a project led by Caltech, funded by NASA, and managed by JPL, and data from the \hxmt mission, a project funded by the China National Space Administration (CNSA) and the Chinese Academy of Sciences (CAS). This research also has made use of the \nustar Data Analysis Software (NUSTARDAS) jointly developed by the ASI Science Data Centre (ASDC, Italy) and Caltech. 

We are grateful to the \hxmt, \srg, and \nustar teams for approving and rapid scheduling of the monitoring campaign. 
\end{acknowledgements}

\bibliography{allbib}
\bibliographystyle{aa}

\end{document}